\documentclass[reprint,pra,showpacs,preprintnumbers,amsmath,amssymb,aps,superscriptaddress]{revtex4-1}
\usepackage{graphicx}
\usepackage{grffile}
\usepackage{tikz}
\usepackage{dcolumn}
\usepackage{amsmath}
\usepackage{times}
\usepackage[braket, qm]{qcircuit}
\usepackage{physics}
\usepackage{mathtools}
\usepackage{layouts}
\usepackage{wasysym}
\begin{document}

\title{Critical faults of leakage errors on the surface code}

\author{Natalie C. Brown}
\affiliation{School of Physics,
Georgia Institute of Technology, Atlanta, GA, USA}
\author{Andrew W. Cross}
\affiliation{IBM T. J. Watson Research Center, Yorktown Heights, NY, USA}
\author{Kenneth R. Brown}
\affiliation{School of Physics,
Georgia Institute of Technology, Atlanta, GA, USA}
\affiliation{Schools of Chemistry and Biochemistry 
and Computational Science and Engineering, Georgia Institute of Technology, Atlanta, GA, USA}
\affiliation{Departments of Electrical and Computer Engineering, Chemistry and Physics, Duke University, Durham, NC, USA }

\date{\today}

\begin{abstract}
Leakage is a particularly damaging error that occurs when a qubit leaves the defined computational subspace. Leakage errors limit the effectiveness of quantum error correcting codes by spreading additional errors to other qubits and corrupting syndrome measurements. The effects of leakage errors on the surface code has been studied in various contexts. However, the effects of a leaked data qubit versus a leaked ancilla qubit can be quite different.  Here, we study the effects of data leakage and ancilla leakage separately. We show that data leakage is much less damaging. %We also show that the distance damaging fault in the surface code comes from ancilla leakage at a particular point in the syndrome extraction circuit. Finally, we simulate two physical realizations of these toy models. 
We show that the surface code maintains its distance in the presence of leakage by either confining leakage to data qubits or eliminating ancilla qubit leakage at the critical fault location. 
We also introduce new techniques for handling leakage by using gates with one-sided leakage and by mixing two types of leakage reducing circuits: one to handle data leakage and one to handle ancilla leakage. 

%We further show techniques for maintaining fault tolerance by mixing two types of LRCs: one to handle data leakage and one to handle ancilla leakage. 

%If leakage errors can be confined to data qubits, and efficiently removed with a minimum overhead leakage reducing circuit (LRC), then the surface code maintains its effective code distance. We also show that the distance damaging fault in the surface code comes from ancilla leakage at a particular point in the syndrome extraction circuit. 
%If leakage errors are eliminated in this particular part of the circuit, the surface code can maintain its effective distance, regardless of other leakage errors that occur on either ancilla or data qubits.Finally, we simulate two physical realizations of these toy models that can be applied to both superconducting and ion trapped architectures. We show that the surface code maintains its distance in the presence of leakage by either confined leakage to data qubits or eliminating leakage at the critical fault location. We further show techniques for maintaining fault tolerance by mixing two types of LRCs: one to handle data leakage and one to handle ancilla leakage. 

\end{abstract}
\maketitle
 \section{Introduction}
 \label{intro}
Qubits are defined as two level systems. Quantum computation relies on the state of the qubit being either in the computational states $\ket{0}$ or $\ket{1}$, or some superposition of both. Unfortunately, most of the devices we build qubits from are not isolated two level systems. They possess states outside of the defined computational basis. When the state of the qubit moves beyond the defined computational states, we say the qubit has leaked. A leaked qubit must be reset for computation to continue. Leakage differs from erasure errors or qubit loss by being undetectable (the locations of leakage errors are unknown) and so it is far more damaging \cite{stace2009thresholds, fujii2012error, barrett2010fault}. Leakage errors are inherent to many qubit architectures including trapped ions \cite{duan2001geometric, haffner2008quantum, cirac1995quantum, plenio1997decoherence, bruzewicz2019trapped, negnevitsky2018repeated, ge2019trapped, bermudez2019fault}, quantum dots \cite{byrd2005universal, fong2011universal, mehl2015fault,andrews2019quantifying, Cai}, superconducting qubits \cite{zhou2005rapid, motzoi2009simple, ferron2010intrinsic, herrera2013tradeoff, ghosh2013understanding, wood2018quantification, bultink2019protecting, varbanov2020leakage} and anyons \cite{xu2008constructing, ainsworth2011topological}. 

Topological surface codes are a leading candidate for handling errors that occur during a computation. They have high thresholds, require only nearest neighbor interactions, and have efficient decoders \cite{fowler2012topological, dennis2002topological, trout2018simulating, fowler2012towards, bravyi2010tradeoffs, calderbank1996good}. However, surface codes can only handle errors within the computational subspace. Leakage errors are extremely damaging to quantum error correcting codes. Leakage errors not only spread additional errors to other qubits, but also lead to measurement errors and will continue to accumulate unless removed. Fortunately a threshold exists for the surface code in the presence of leakage errors if extra circuits are used to return leaked qubits to the computational space \cite{aliferis2005fault}. 

Leakage reducing circuits (LRCs) remove leakage from the system by swapping reinitialized qubits for leaked ones \cite{aliferis2005fault, suchara, fowler, ghosh2015leakage}. While this is an effective means of removing leakage, most LRCs have a substantial overhead. Previous work has shown that implementing LRCs after every gate, for both ancilla and data qubits in the surface code, is a fault-tolerant way of handling leakage errors. But incorporating more resource efficient LRCs results in an effective distance suppression due to the presence of leakage \cite{suchara}. Because of this, and the expense associated with implementing these circuits, understanding exactly when and where these distance damaging leakage faults occur in the surface code is crucial for employing LRCs in an effective manner while minimized overhead. 

There are many different physical mechanisms that can induce leakage but the models describing how leaked qubits interact with other qubits fall into two fundamental categories: interacting and non-interacting. In an interacting leakage model, when a leaked qubit is involved in a gate with an unleaked qubit, the two qubit interaction induces some error on the unleaked qubit. The worst case scenario of this is if the leaked qubit actually induces a leakage error on the unleaked qubit. This can be seen in silicon based architectures \cite{Cai}.  A slightly better, albeit still damaging, leakage interaction model is know as the depolarizing leakage model and is more commonly studied in ion trap and superconducting architectures. In this model the leaked qubit completely depolarizes the unleaked qubit. It is well known that this leakage model results in a suppression of the code distance of the surface code \cite{fowler, suchara, brown2018comparing, mike, mixed}. 

In a non-interacting leakage model, the two qubit interaction does not occur when one or both inputs have leaked and thus does not change the state of the unleaked qubit. However, because most naive gate decompositions are made up of both single and two qubit gates and there is no way of knowing a qubit leaked mid-circuit, single qubit gates are still applied and this can change the state of the unleaked qubit. Non-interacting leakage models are highly dependent on the gate implementation and the physical device. Because they tend to have a bit more structure than interacting leakage models, non-interacting leakage models tend to not be as damaging on quantum error correcting codes \cite{mike, mixed}.

%In addition to the different models for leakage interaction, leakage errors themselves have two base models: detectable and undetectable. Detectable leakage, also known as erasure errors, 
In this work, we focus on the depolarizing leakage model in an effort to understand how to fault-tolerantly handle leakage of this nature. We first study ancilla leakage and data leakage separately to understand the effects leakage has on each type of qubit. We then isolate leakage to certain parts of the circuit in an effort to analyze where the critical leakage faults lie. Finally, motivated by our observations, we construct different fault-tolerant schemes for handling leakage on the surface code that can be applied to both superconducting and ion trapped architectures. 
%We end by make a short observation on the irrelavence of how leakage is measured. 

\section{Topological surface codes and Leakage reducing circuits}
 
 \begin{figure}[ht]
%trim={<left> <lower> <right> <upper>}
\includegraphics[trim=0 0 0 0, clip, width=\columnwidth]{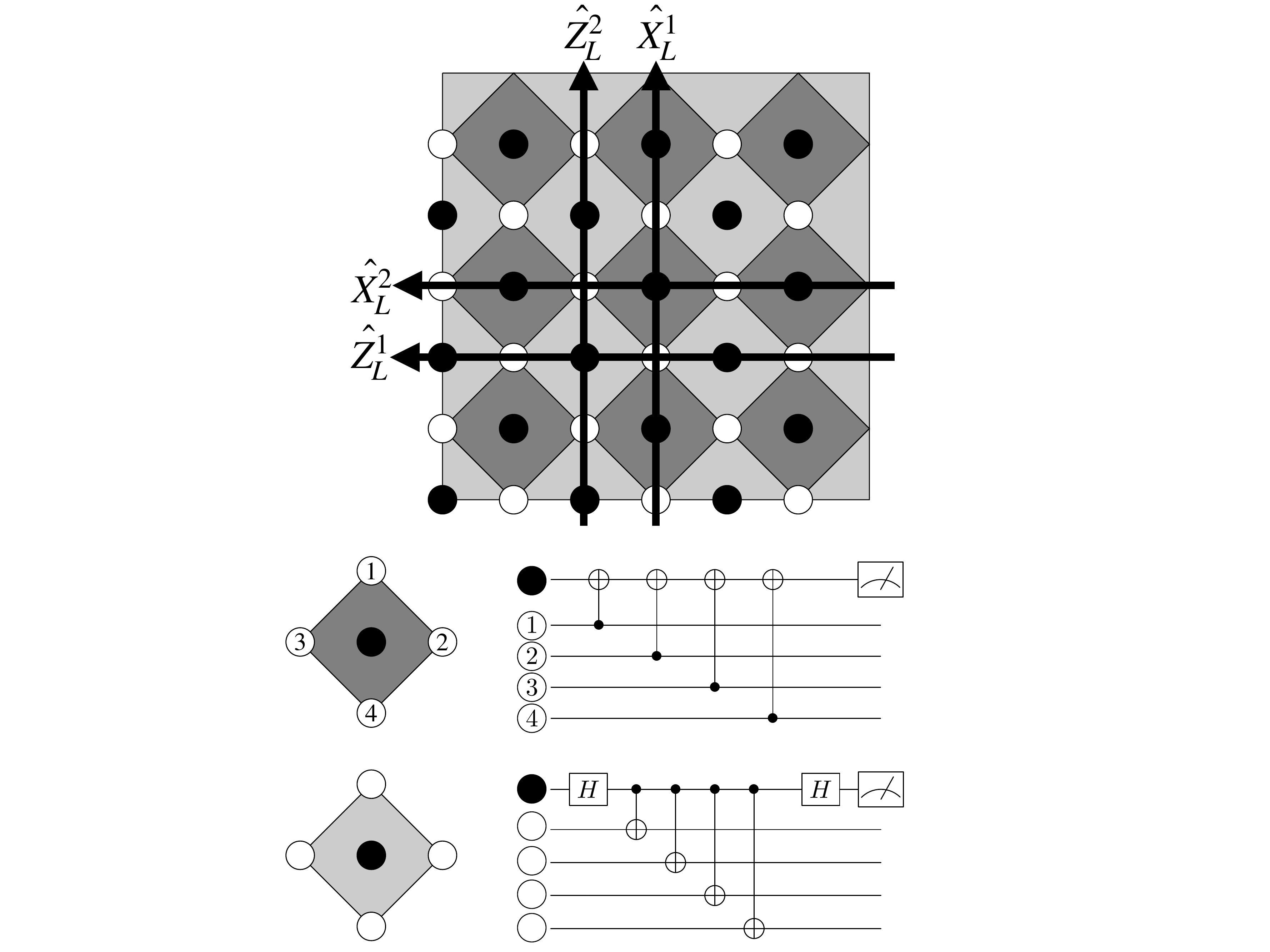}
\caption{Standard circuits to measure toric code check operators. The open white circles represent data qubits while the closed dark circles represent measure/ancilla qubits. The dark and light gray represent $\hat{Z}$ and $\hat{X}$ stabilizers respectively. The toric code encodes two logical operators $\hat{X^1_L}/\hat{Z^1_L}$ and $\hat{X^2_L}/\hat{Z^2_L}$ that span the boundaries of the surface. }
\label{surf}
\end{figure}
Topological quantum error correction codes, such as the surface code, encode information into a topological surface comprised of many physical qubits. The information is stored in the topological degrees of freedom and is thus protected from small local errors that occur on the individual physical qubits \cite{bombin, fowler2012surface, dennis2002topological,raussendorf2007fault, kitaev1997quantum}. A surface code of distance $d$, should be able to detect and correct $\lfloor (d-1)/2 \rfloor$ physical errors. The aim of such codes is to achieve a logical error rate which is \textit{smaller} then the physical error rate. A surface code is said to be fault-tolerant to $\frac{d-1}{2}$ physical errors, if the scaling of the logical error rate obeys the power law
 \begin{equation}
P_L \propto   P^{\lceil\frac{ d}{2}\rceil}
\label{maintain}
 \end{equation}
where $P_L$ is the probability of a logical error, $P$ is the probability of a physical error, and $d$ is the code distance \cite{fowler2012surface}. 
 
In the depolarizing leakage model, a single leakage error can produce a two qubit error chain. The surface code can then only be guaranteed to correct $\lfloor (d-1)/4 \rfloor$ physical errors and thus the codes effective distance is halved. In this case the logical error rate is reduced to
\begin{equation}
P_L \propto P^{\lceil\frac{ d}{4}\rceil}
\label{drop}
\end{equation}
This suppression of the code distance is a well known effect of leakage on the surface code \cite{fowler, suchara, brown2018comparing, mike, mixed}. 

In the surface code, qubits function either as data qubits, which hold the encoded information, or ancilla qubits, which are used to detect errors. Each ancilla qubit is used to measure either a $\hat{X}$ or a $\hat{Z}$ type stabilizer. The process of measuring the stabilizers is completed in a six step cycle. First ancilla qubits are initialized into either the $\ket{0}$ ($\hat{Z}$ stabilizers) or $\ket{+}$ ($\hat{X}$ stabilizers) state. Then a series of 4 CNOTs are performed between the ancilla qubits and its neighboring data qubits. Finally, the ancilla qubits are measured in the respective basis \cite{suchara, fowler2012topological, kitaev2002classical}. Fig. \ref{surf} shows a distance 3 toric code and the stabilizer measurement circuits, often referred to as syndrome extraction circuits. The toric code encodes two logical qubits into $d^2$ physical qubits and requires an additional $d^2$ qubits to check for errors. 

Surface codes are not designed to detect or correct leakage errors. Leakage must be actively removed from the code or else it will eventually saturate the entire system. At the logical level, this is achieved by incorporating LRCs which convert leakage errors into Pauli-type errors which can be detected and corrected by the code \cite{aliferis2005fault}. 

There a few different proposed LRCs \cite{aliferis2005fault, ghosh2015leakage, suchara}, but most rely on the same technique of swapping in newly prepared qubits for leaked ones. The main difference among most proposed LRCs is the amount of swapping done and overhead of additional qubits. For example, swapping new qubits to both ancilla and data qubits after every gate is known as the Full LRC \cite{suchara} and requires double the amount of qubits. Swapping only at the end of the syndrome extraction cycle to only the data qubits is known as the circuit LRC \cite{suchara} and requires the addition of half the number of qubits. For an overview and comparison of the different overheads between these circuits, we recommend this paper \cite{suchara}.
 
\begin{figure}[ht]
%trim={<left> <lower> <right> <upper>}
\includegraphics[trim=0 280 0 240, clip,width=\columnwidth]{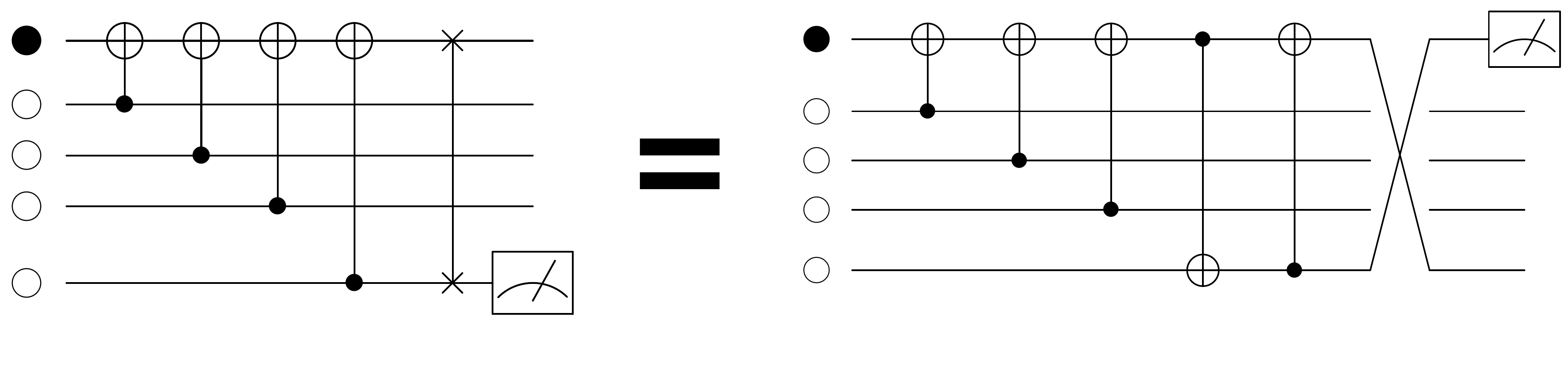}
\caption{The SWAP LRC changes the roles of data and ancilla qubits every round of syndrome extraction. Leakage is removed from the system through measurement and reset of ancilla qubits. This minimum overhead LRC add a single additional gate to the circuit.}
\label{swap}
\end{figure}

 Of particular interest to us, is the "QUICK" or "SWAP" LRC \cite{brown2018comparing, mike, ghosh2015leakage, mixed}. This LRC swaps data qubits with ancilla qubits at the end of the circuit and thus requires no additional qubits to implement. Adding the SWAP gate amounts to adding one additional gate to our syndrome extraction circuit (see Fig. \ref{swap}). Ancilla qubits get reinitialized every round of syndrome extraction so swapping data qubits for ancilla qubits removes leaked qubits from the system. Qubits switch their roles (data/ancilla), every round of syndrome measurement. Of circuit level leakage reduction techniques, it has the lowest overhead. 

\section{Ancilla leakage vs Data leakage}
\begin{figure}[ht]
\includegraphics[width=\columnwidth]{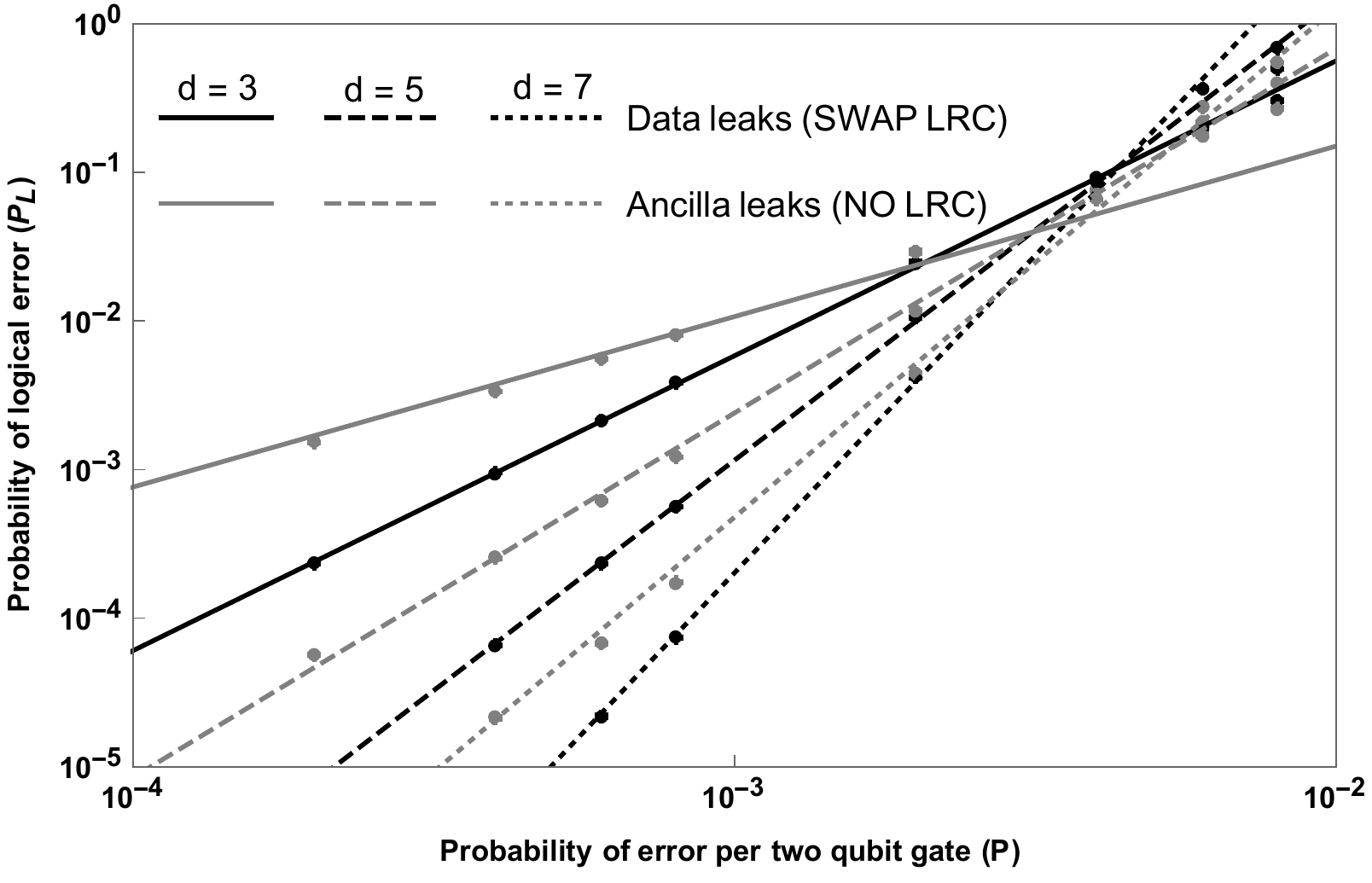}
\caption{Comparison of the logical error rate of the surface code. When leakage errors are confined to data and removed using a SWAP LRC, there is a substantial gain in the logical error rate compared to when leakage errors are confined to ancilla only. Note the incorporation of the SWAP LRC used to handle data leakage requires one extra gate.}
\label{AvsD}
\end{figure}

 Motivated by our results of mixing qubits types on the surface code \cite{mixed}, we decided to investigate the effects of data leakage and ancilla leakage separately. If leakage errors are limited to ancilla qubits only, there is no need to implement a SWAP LRC. In fact, implementing a SWAP LRC will only allow leakage errors to live longer and induce more errors into the system due to the extra fault locations associated with the extra gate. Data qubits are never reset and require the use of an LRC. We simulated two different systems: one where only ancilla leak and no LRC is needed, and one where only data leak so a SWAP LRC is used adding additional fault locations. The results of these simulations can be seen in Fig. \ref{AvsD}.

It is clear from these results, ancilla leakage is much more damaging than data leakage. We observed the code distance suppression characteristic of leakage errors when leakage is confined to the ancilla qubits (i.e. obeys Eq. \ref{drop}) but our effective distance is maintained when leakage is confined to the data qubits with an LRC (i.e. obeys Eq. \ref{maintain}). 

\subsection{Propagation of errors from data leakage}
\begin{figure*}
\includegraphics[trim={0 175 0 80}, clip,width=\textwidth]{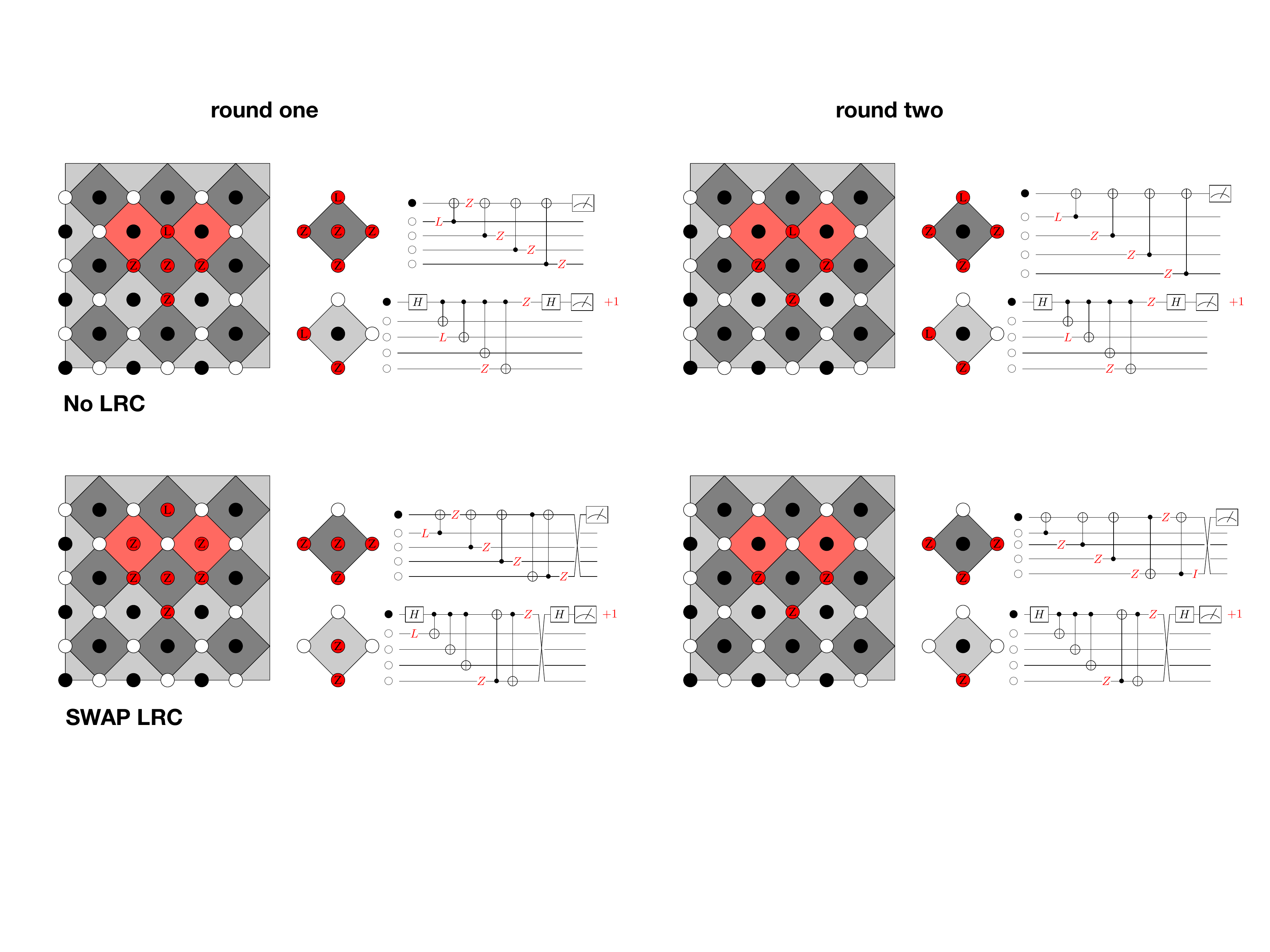}
\caption{An example of the spread of errors caused by data leakage. In both cases we consider a leakage error on a data qubit, that spreads a $\hat{Z}$ error to an ancilla qubit. This $\hat{Z}$ error than propagates on to additional data qubits. In order to understand how the SWAP LRC handles errors of this nature, we need to look at multiple rounds of syndrome extraction. When no LRC is used (top), a single leakage error on a data qubit, can spread 3 or more errors on to additional data qubits.
Since the leakage error is not removed, it can further spread errors subsequent rounds of syndrome extraction. If the SWAP LRC is used (bottom), then the leakage error is removed and the resulting error configuration is equivalent to a single physical error. }
\label{DataLeakage}
\end{figure*}
Data leakage is fundamentally bad. If leaked data qubits are not reset in some way, the leakage will accumulate until all information is lost. Consider the error configuration of a distance 3 toric code in Fig. \ref{DataLeakage}. A single leakage error on data qubit induced a $\hat{Z}$ error on to a $\hat{Z}$ stabilizer ancilla qubit. This ancilla qubit then spreads additional $\hat{Z}$ errors to other data qubits. This single leakage error has now produced 3 additional errors. Note that if this was a Pauli error, this error propagation would not happen. 

In order for the surface code to be fault-tolerant against measurement errors, we perform multiple rounds of syndrome extraction before we decode. Without the use of the SWAP LRC, leakage errors are not removed and can potentially spread additional errors. When the stabilizers are measured, the syndrome information will indicate on error occurred at the on the qubit that leaked. However, applying a correction to the leaked qubit will not correct the leakage. Unless the leakage error decays back to the computational subspace, or is removed, we will continue to get compromised syndrome information. By adding the SWAP gate at the end, we not only removed the leakage error, we now are able to apply the proper correction based on the syndrome information.

 %However, the surface code is relatively robust to data leakage. If data leakage is properly removed, then the type of errors it can produce are all measurement errors. Data leakage can spread 1, 2, 3, or 4 errors to ancilla. Since it will randomly spread these errors every syndrome extraction cycle, and the qubits it spreads errors to are ancilla that get reset every round, the amount of damage it can cause it limited to measurement errors.
 
\begin{figure}[ht]
\includegraphics[width=\columnwidth]{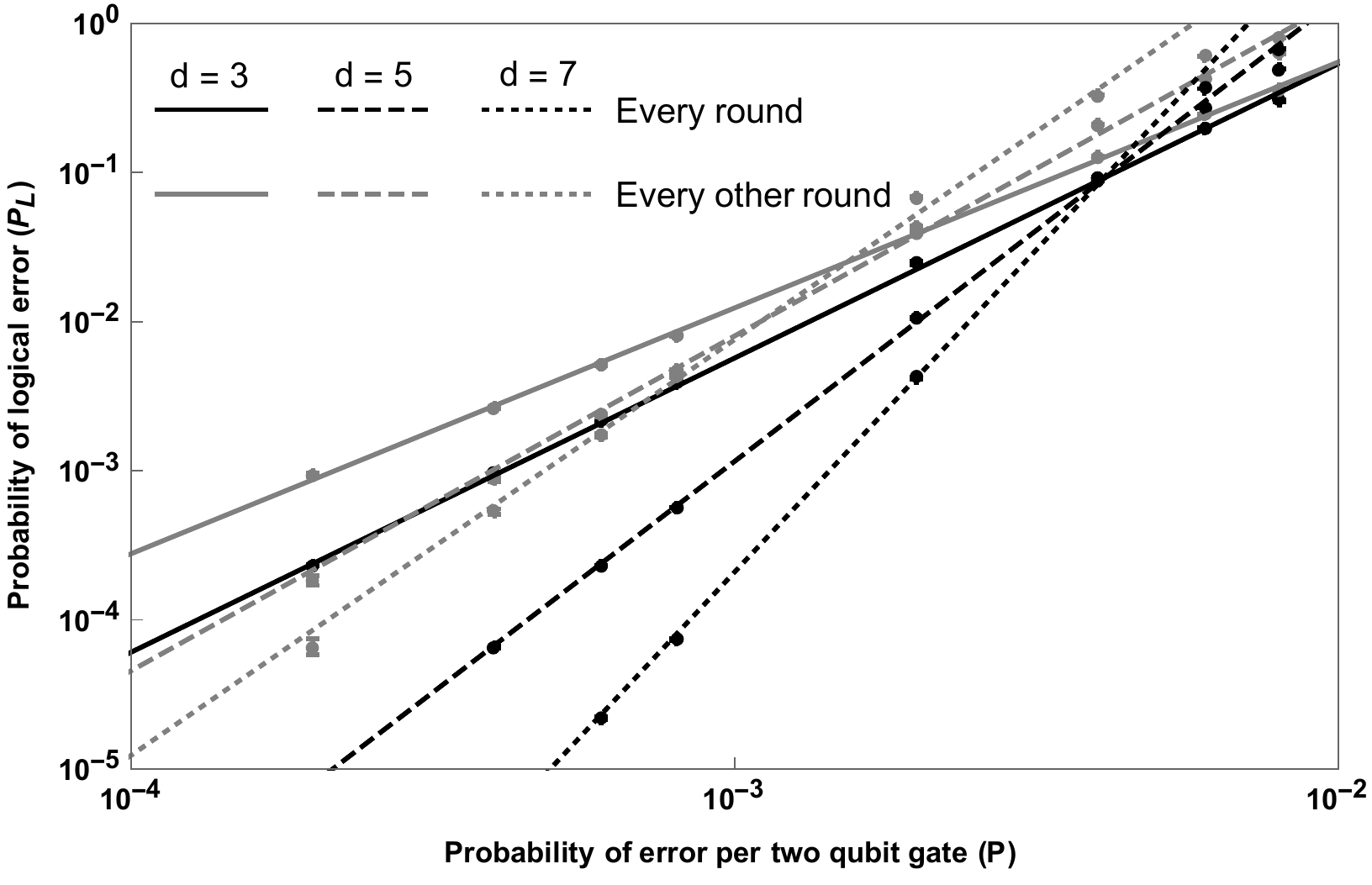}
\caption{A comparison of the surface using alternating syndrome extraction circuits. Either the SWAP LRC is implemented every round (black) or every other round (gray). The SWAP LRC must be implemented every round of syndrome extraction to effectively mitigate the errors spreading from the leakage error.}
\label{DataRemove}
\end{figure}

 In order to test the robustness of the surface code to data leakage, we ran simulations implementing the SWAP LRC at various periods of syndromes extraction. The results of Fig. \ref{DataRemove} show implementing the SWAP LRC at every round of syndrome extraction, meaning the circuit in Fig. \ref{swap} was used every time, and every other round, meaning the simulation alternated between the circuits in Fig. \ref{surf} and Fig. \ref{swap}. These results show the importance of removing leakage from the system efficiently and how the effectiveness of the SWAP LRC is dependent on successive implementation. 
Putting this together with our results from Fig. \ref{AvsD}, we conclude that if leakage is confined to data qubits, and removed every round of syndrome extraction, then the effective distance of the surface code can be maintained. 

 \subsection{Propagation of errors from ancilla leakage}
\begin{figure}
%trim={<left> <lower> <right> <upper>}
\includegraphics[trim=100 460 0 135, clip, width=\columnwidth]{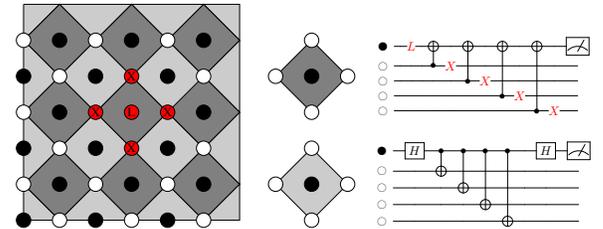}
\caption{Leakage errors on ancilla are particularly harmful as they can spread hook errors. The worst case scenario is seen here, where a single leakage error spread to 4 physical errors.}
\label{ancillaLeakage}
\end{figure}
Naively, ancilla leakage does not seem like it would be as damaging as data leakage since ancilla qubits get reinitialized often. Ancilla leakage will spread errors to data qubits. The single qubit errors it can spread will look like measurement errors and will not be a problem, as they will be detected and corrected by multiple rounds of syndrome extraction. Two qubit errors come in two different flavors, benign two qubit errors that propagate perpendicular to the logical operator, and malignant two qubit hook errors that run parallel to the logical operator. 
 
Consider the error configuration in Fig. \ref{ancillaLeakage}. A single leaked ancilla after the first CNOT gate spreads four $\hat{X}$ errors running along the support of each logical operator (see Fig. \ref{surf}). Since the toric code encodes two logical qubits, there are two different hook errors that can occur. We shall see in section \ref{criloc}, that this is why initialization leakage is more damaging than leakage at other faults. 
 
In a rotated surface code, it is important to order the gates in a way that Pauli errors propagate perpendicular to the direction of the logical operator \cite{tomita, trout2018simulating}. This is then equivalent to a single-qubit error maintaining fault-tolerance with bare ancilla. In unrotated codes, like the unrotated toric code considered here, this scheduling becomes less important for Pauli errors. However because leakage errors cause depolarizing noise, clever gate scheduling like that found in \cite{tomita}, can minimize harmful hook errors. Unfortunately, there is no scheduling that can completely eliminate these hooks.

The results of Fig. \ref{AvsD} show that this suppression of the code distance is caused by these hook errors originating from a leaked ancilla qubit. This knowledge motivates the design of circuits that can utilize these facts and confine leakage errors to data qubits.

\section{Confining leakage to data}
While other mechanisms such as initialization can cause leakage, the predominate source of leakage in the surface code arises from the single and two qubit gates in the syndrome extraction circuit. Many physical processes that cause leakage have no bias to which qubit can leak in a two qubit gate, such as spontaneous scattering from Raman driven gates in ion-trapped devices \cite{ozeri2005hyperfine, brown2018comparing}. However, there are a few examples when the structure of the qubit and implementation of the gate produce a biased in the direction of leakage. We call this a one-sided leakage model; one qubit in a two qubit gate is much more likely to leak than the other. 

One example of a one-sided leakage model can be found in the cross-resonance (CR) gate used for transmon qubits \cite{magesan2018effective, kirchhoff2018optimized}. In CR gate, the control is driven at the target qubit frequency. This induces Rabi oscillations of the target qubit with a frequency depending on the state of the control qubit. This entangles the two qubits providing an operation locally equivalent to a CNOT. Because only the control is driven, the control has a much higher probability to leak compared to the target \cite{PhysRevA.100.012301, wood2018quantification}. 
%%%%%%%%%NOTES%%%%%%%%%%%

\begin{figure}
%trim={<left> <lower> <right> <upper>}
\includegraphics[trim=90 0 100 0, clip, width=\columnwidth]{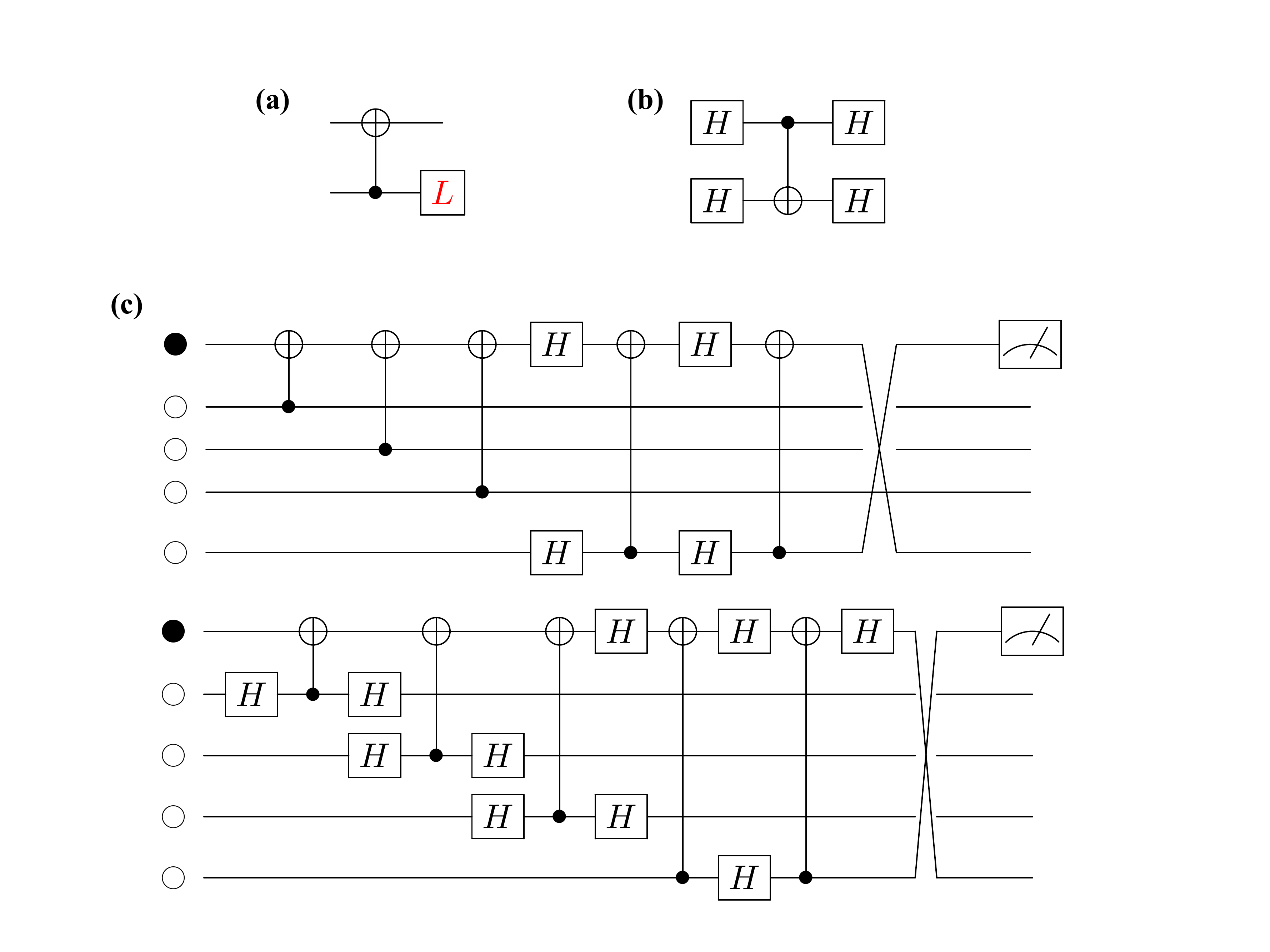}
\caption{(a) In a one-sided leakage model, leakage errors only occur on one of the qubits involved in a CNOT gate. (b) The addition of 4 single qubit $H$ gates, reverses the direction of the CNOT gate. (c) The gate biased circuit uses 12 additional single qubit gates to confine leakage errors to data qubits. Here the top circuit is the $\hat{Z}$ type syndrome extraction circuit and the bottom is the $\hat{X}$ type syndrome extraction circuit. }
\label{gb_iden}
\end{figure}

This physical realization of the one-side leakage model encourages us to make use of gate identities and design syndrome extraction circuits that confine leakage to the data. The direction of the CNOT gate can be easily changed by conjugating it by $H$ gates. The addition of 12 single qubit $H$ gates isolates leakage errors to the data, which are then easily removed with the SWAP LRC (see Fig. \ref{gb_iden}). We call this circuit the gate biased circuit. Note that this is already better than the standard circuit because leakage is only allowed to live for one round of syndrome extraction. 

\begin{figure}[ht]
\includegraphics[width=\columnwidth]{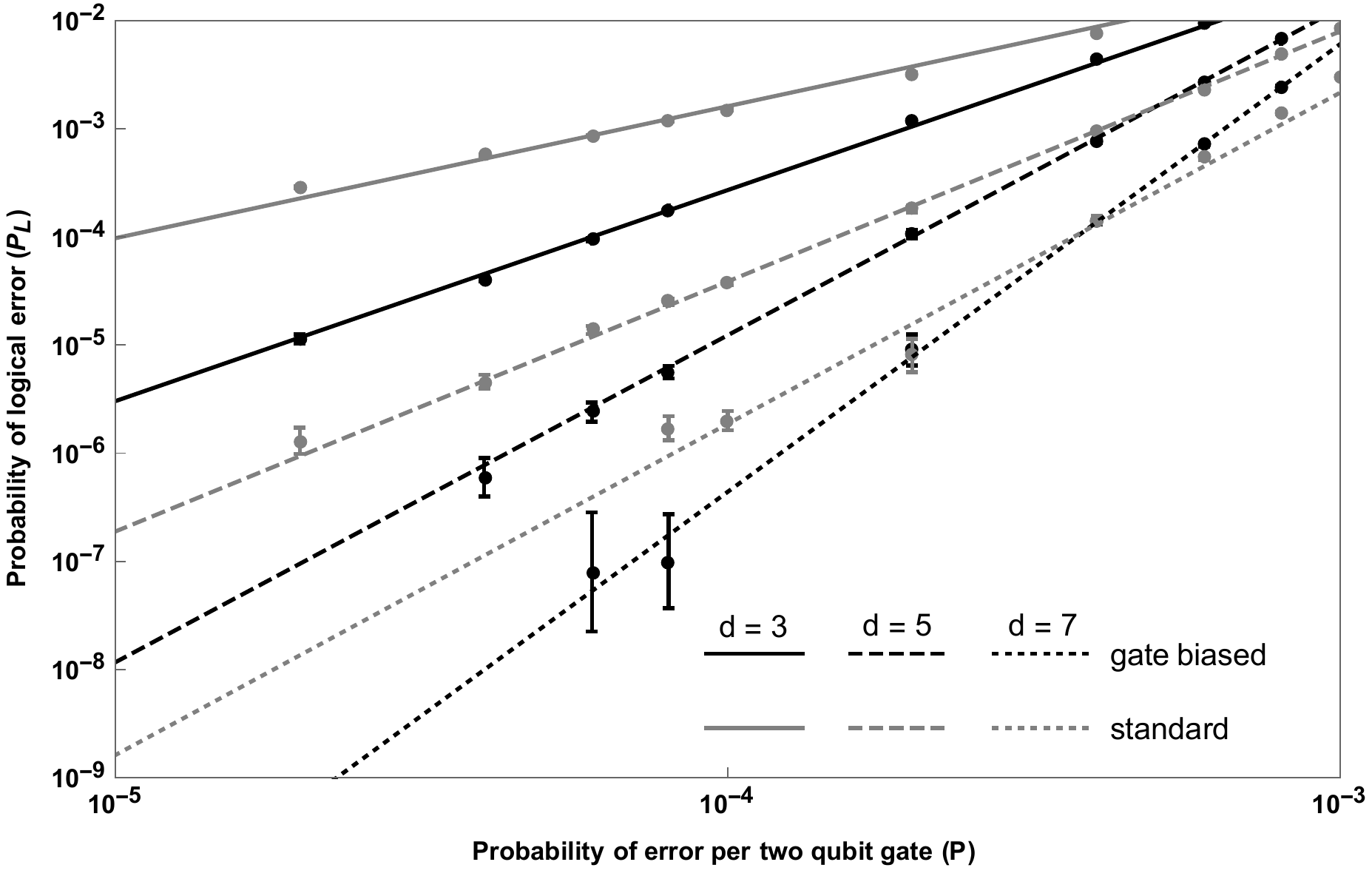}
\caption{A comparison of the logical error rate of the surface code between the standard syndrome extraction circuit (gray) and the gate-biased model (black). Here the leakage to depolarizing gate noise was 1:1. The gate-biased model isolated leakage events to the data qubits at the cost of requiring 12 additional single qubit gates.}
\label{gbtrans}
\end{figure}

We ran simulations comparing the standard circuit, in which no single qubit gates are added, and the gate biased circuit. It is important to note that in both simulations, only the control leaks. However, because there are controls on both data and ancilla qubits in the standard circuit, the standard circuit is a two-sided leakage model. We also included single qubit gate leakage. 
The results can be seen in Fig. \ref{gbtrans}. In our simulation, initialization did not cause leakage. We shall discuss initialization leakage separately. Furthermore leakage was assumed to be undetectable. The ratio of leakage to depolarizing gate noise was 1:1. %These rates are motivated by superconducting and ion-trap devices respectively.

%\begin{figure}[h]
%\includegraphics[width=8cm, height=5cm]{figures/GB_ion.pdf}
%\caption{gate biased circuit, ion}
%\label{gbion}
%\end{figure}

The gate biased circuit outperforms the standard circuit. Adding 12 single qubit gates and the extra errors associated with those gates, effectively isolates leakage to the data and the code distance is preserved. Leakage faults occurring on the ancilla in the standard circuit suppress the code's effective distance.

\section{Critical leakage locations}
\label{criloc}
While the gate biased circuit offers a clear advantage over the standard circuit, the addition of 12 single qubit gates adversely effects our overall error rate. Knowing these critical leakage faults originate from ancilla (if data leakage is efficiently removed), we can further isolate where these critical faults lie by isolating leakage events in the circuit. 

%We looked at two different scenarios: one where only one CNOT gate allowed leakage, and one where only one CNOT gate does \textit{not} allow leakage. These simulations assumed a two-sided leakage model (i.e. both data and ancilla can leak) and implement the SWAP LRC. In both simulations for simplicity, we again did not include initialization leakage. The results of these simulations agreed with each other. 

\begin{figure}
\includegraphics[width=\columnwidth]{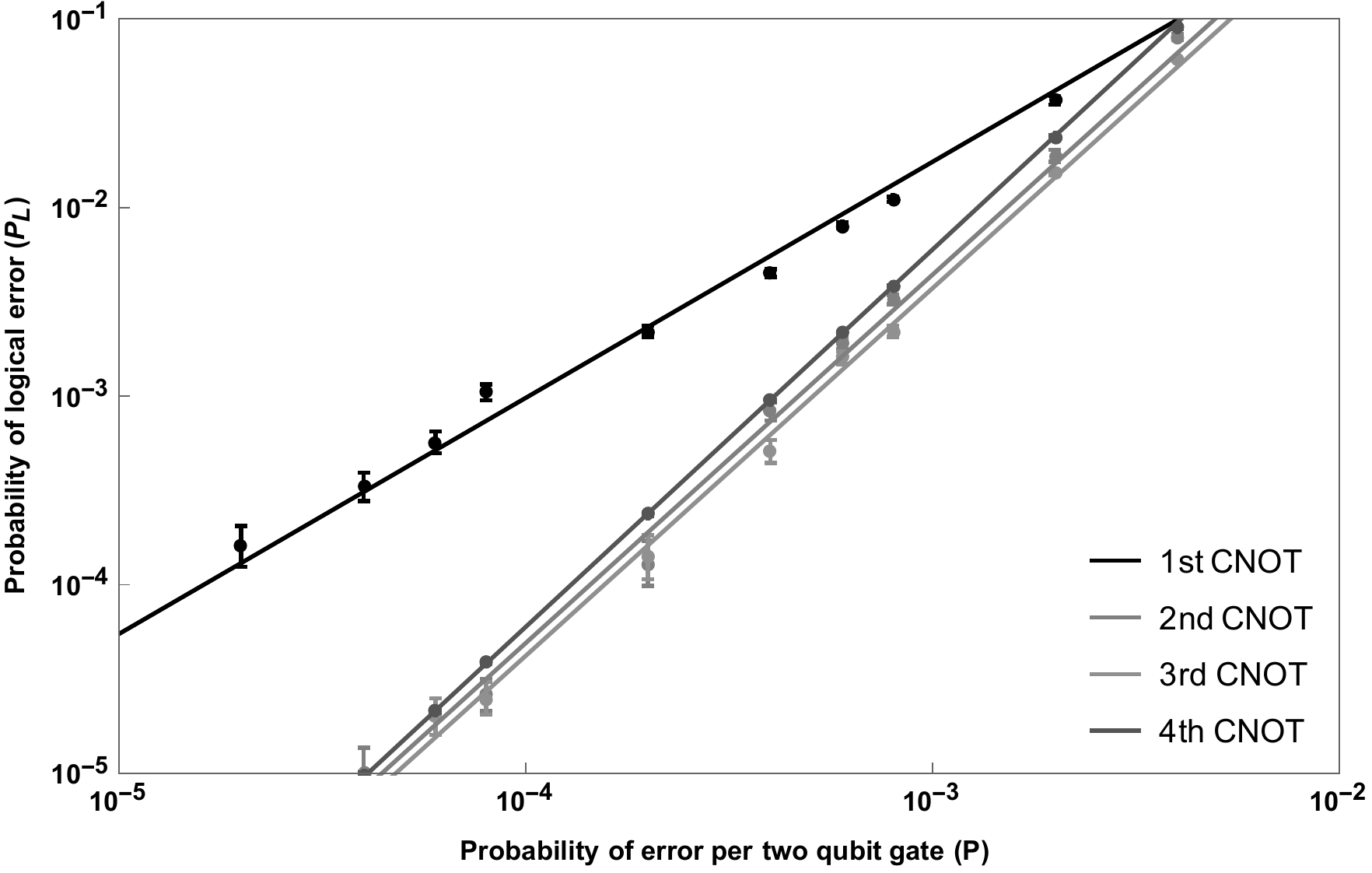}
\caption{Results of confining leakage errors to one gate in a distance 3 toric code, with a SWAP LRC. By isolated leakage events we can identify where the critical fault locations lie. In these simulations, only one CNOT caused leakage. It is clear that there is a significant improvement in the logical error rate if we do not allow leakage from the 1st CNOT. In these simulations, we did not allow for initialization leakage.}
\label{OL3}
\end{figure}

Figure \ref{OL3} shows the logical error rate when only one two qubit gate in the syndrome extraction circuit can cause leakage. Of particular note, if leakage is isolated to the 2nd, 3rd, or 4th CNOT, than the effective code distance is maintained. Only leakage events from the 1st CNOT gates produce the hook errors and lead to the distance suppression. This is a direct result from our gate scheduling. Leakage events anytime after the 2nd CNOT will cause 2 additional physical errors at most, along the same support of the stabilizer. This simulation assumed a two-sided leakage model (i.e. both data and ancilla can leak) and implement the SWAP LRC.

In order to simplify the simulation, we did not include initialization leakage. Initialization leakage would allow for the same hook error as leakage errors from the 1st CNOT. Initialization leakage errors are the most damaging as there are more possible combinations of hook errors they can produce. 

The gate scheduling minimizes the critical leakage fault locations to after initialization and the 1st CNOT gate. It is important to note that the first $H$ in the $X$ stabilizer syndrome extraction circuit also is a critical fault location and was included in all these simulations. However, focusing efforts on minimizing leakage from initialization and the 1st CNOT will suffice for taking care of this additional critical fault location.  

\subsection{Optimized gate biased circuit}
Identifying these critical leakage fault locations allows us to optimized the gate biased circuit. Since leakage need only be eliminated at the early part of the circuit, we can reduce the number of additional single qubit gates from 12 to 4. 

\begin{figure}
%trim={<left> <lower> <right> <upper>}
\includegraphics[trim=150 395 150 150 clip, width=\columnwidth]{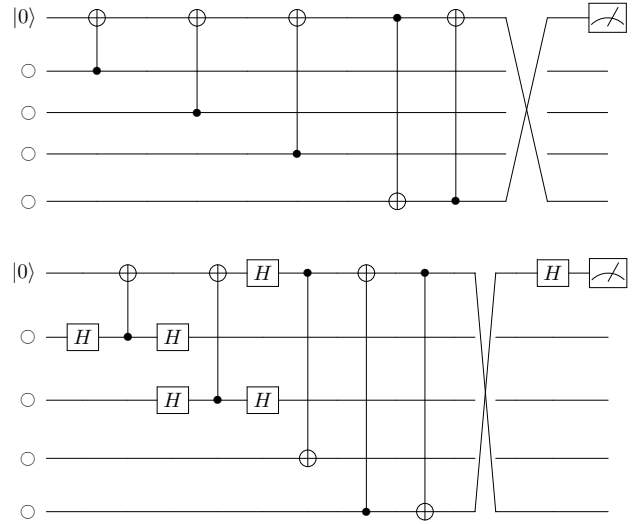}
\caption{Knowing that the worst errors come from ancilla leakage at the beginning of the circuit, we can make an optimized model. Instead of implementing 12 additional $H$ gates and completely isolated leakage to data qubits, we need only isolated leakage errors at the very beginning of the circuit. This reduces the number of single qubit gates needed from 12 to 4.}
\label{optgbmodel}
\end{figure}

Fig. \ref{optgbmodel} shows the optimized circuits. Naively, one might think it is good enough to only change the direction of the 1st CNOT gate, since ancilla leakage from the 2nd CNOT is not a critical fault. However, adding the additional single qubit gates required, leaves the circuit open to single qubit gate leakage at the critical fault location. Flipping the 2nd CNOT gate gives us a few more single qubit gates which cancel and eliminate leakage at the critical fault location. It is important to note, this scheme will not work with initialization leakage, as it does not eliminate leakage at the initialization critical fault, however many initialization schemes do not cause leakage \cite{olmschenk2007manipulation, PhysRevA.82.063419, noek2013high}. We shall discuss methods for handling initialization leakage in the next section.   

\begin{figure}[ht]
\includegraphics[width=\columnwidth]{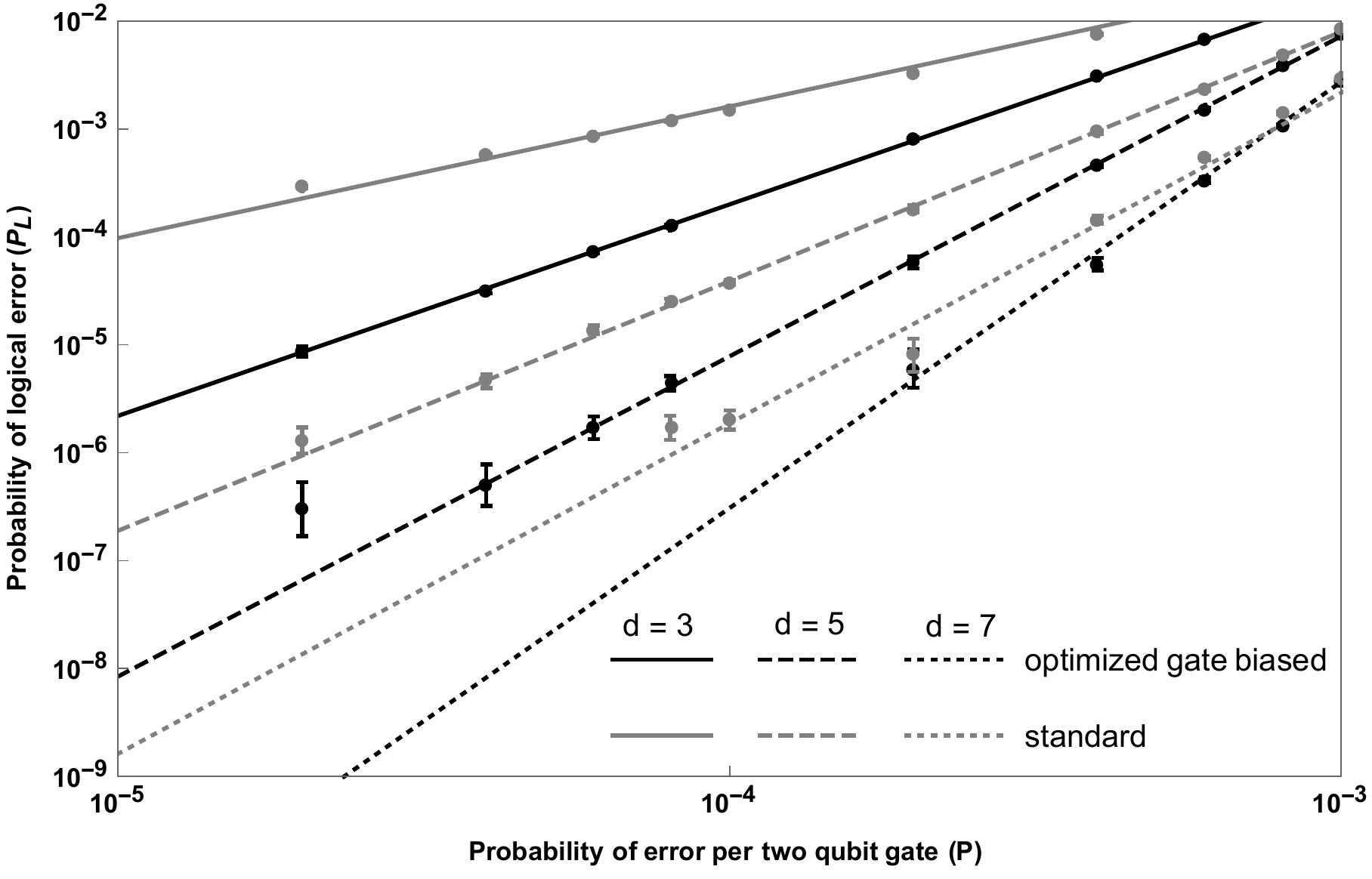}
\caption{A comparison of the logical error rate of the surface code between the standard syndrome extraction circuit (gray) and the gate-biased model (black). Here the leakage to depolarizing gate noise 1:1. The gate-biased model isolated leakage events to the data qubits at the cost of requiring 4 additional single qubit gates.}
\label{opttrans}
\end{figure}

We again test the new optimized circuit against the standard circuit (Fig. \ref{opttrans}). The optimized model effectively maintains the distance by not completely isolating leakage to data, but instead eliminating leakage at the critical fault location. It offers improvement over both the standard and gate-biased model as it utilizes fewer single qubit gates. Its advantage over the gate-biased model is not substantial but is still advantageous due to its lower circuit depth. 

%\begin{figure}[h]
%\includegraphics[width=8cm, height=5cm]{figures/Opt_Ion.pdf}
%\caption{optimized, ion}
%\label{option}
%\end{figure}

\subsection{Mixed LRC}
The gate-biased model and its optimized version offer fault-tolerant solutions for the surface code in the presence of leakage. However, both circuits rely on a one-sided leakage model. Identifying the exact location of the critical leakage faults inform us on fault-tolerant schemes for a two-sided leakage model. 

Knowing that data leakage can be effectively removed with the SWAP LRC, and knowing that leakage needs to be eliminated after initialization and the 1st CNOT gate, we construct a new version of an LRC. By swapping in reinitialized qubits after the 2nd CNOT, we handled the critical fault locations caused by initialization and the 1st CNOT. This handles logical errors cause from leaked ancilla. Adding a SWAP LRC at the end handles leakage errors from data. 

The location of the SWAP gate for the ancilla is crucial for the same reasons as the optimized gate-biased model. Swapping in a new qubit too soon will not eliminate leakage at the critical fault location and could in fact induce a leakage error that was not there before. Swapping the ancilla after the 2nd CNOT gate ensures that leakage errors are being swapped out, and if a new leakage error was introduced, it is not at a critical fault location.  

\begin{figure}
%trim={<left> <lower> <right> <upper>}
\includegraphics[trim=200 550 200 100, clip, width=\columnwidth]{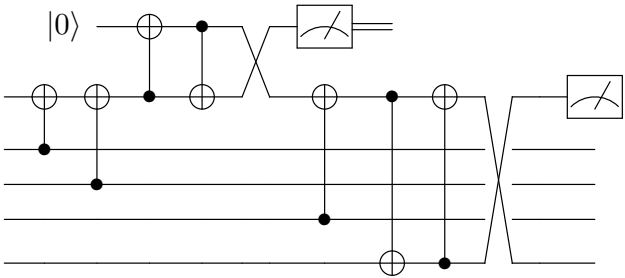}
\caption{By swapping in a new qubit in the middle of the syndrome extraction circuit we can eliminate hook errors that could arise from ancilla leakage. Implementing the SWAP LRC at the end ensures data leakage will not be a problem. This new LRC increases qubit overhead, but is still more economical that other LRCs. }
\label{mixLRCioncir}
\end{figure}

\begin{figure}[ht]
\includegraphics[width=\columnwidth]{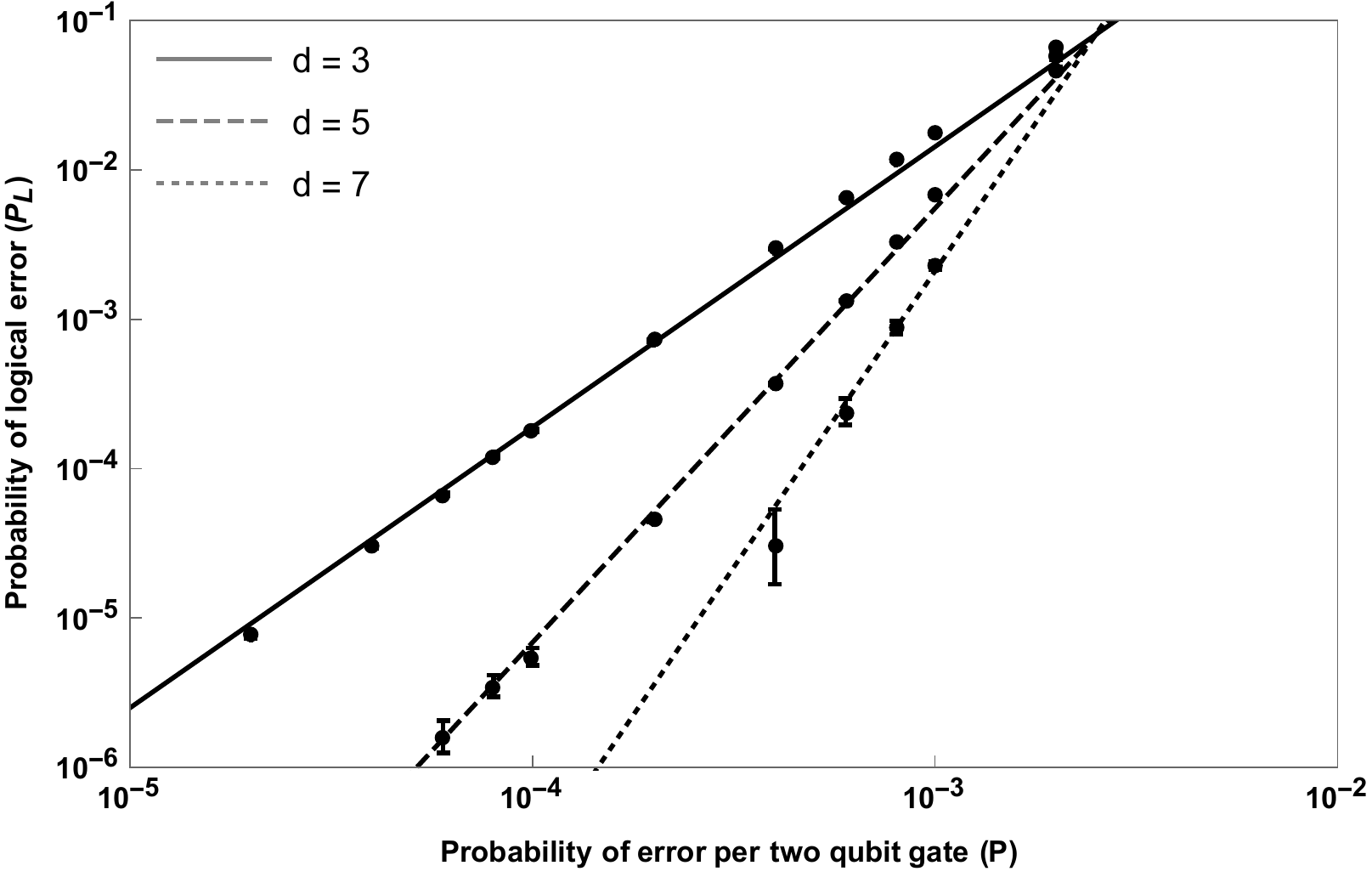}
\caption{The logical error rate of the surface code implementing the mixed LRC. The leakage to depolarizing gate noise ratio is 1:1. The code maintains its effective distance. The mixing of two LRCs handles all both initialization and leakage errors occurring from the first CNOT.}
\label{mixLRCion}
\end{figure}

The probability of a logical error of the surface code implementing this new LRC can be seen in Fig. \ref{mixLRCion}. As before the ratio of leakage to gate noise is 1:1, however this time we allow for initialization leakage. We observe that the effective distance is maintained and that this LRC is an effective means of fault-tolerantly handling leakage. 

To implement this new LRC, the amount of ancilla qubits doubles. However, in terms of qubit count and logical error rate, we believe this is the most efficient fault-tolerant LRC implementation for a two-sided leakage model.  

\section{Conclusions}
Leakage is an exceptionally damaging error. However not all leakage faults are equal. We have shown that ancilla leakage is more damaging that data leakage. A single ancilla leakage error can cause a hook error that will lead to a logical error in a single round of syndrome extraction. A single data leakage error will lead to a logical error if not removed every round of syndrome extraction. If leakage errors can be isolated to data qubits, and efficiently removed every round, the then effective distance of the surface code can be maintained.

Further, a gate scheduling of the syndrome extraction circuits can be done to minimize these hooks, but not to completely eliminate them. With this scheduling \cite{tomita}, two critical fault locations remain: after initialization and after the 1st CNOT gate. If ancilla leakage can be eliminated at these critical faults, and data leakage efficiently removed, then the effective distance of the surface code can be maintained.

Knowing these critical fault locations can help with future circuit designs. A natural extension of this work would be to look into leakage detection \cite{gottesman1997stabilizer, preskill1998reliable} and flag qubit schemes \cite{PhysRevA.101.012342, chao2018quantum} to identify leakage at these faults. This could influence decoding strategies as well. 

It would also be advantageous to investigate different physical mechanisms for leakage elimination \cite{hayes2019eliminating}. Implementing these physical techniques at the critical fault locations would be an effective procedure. Mixing LRC's with physical techniques could help minimized the overhead. 

Finally, we hope that this analysis of leakage faults in the surface code offers some insight as to how to fault-tolerantly handle leakage in other codes, particularly the triangular color code \cite{chao2018fault, bombin2007topological}. While leakage errors will not be a limiting factor in current devices, we hope this work provides some intuition on how to handle leakage in future large scale architectures.

\section{Acknowledgments}
We would like to thank Dripto Debroy, Muyuan Li, and Michael Newman for useful discussions on leakage models and surface codes. N.B would like to thank Michael Foss-Feig, Bryce Bjork, and Dan Stack for insight on leakage in trapped-ions architectures. And finally we would like to thank Jay Gambetta who gave us the idea to investigate the one-sided leakage model and Ted Yoder for useful feedback. N.B. is funded in part by ODNI/IARPA LogiQ program (W911NF-16-1-0082) and also funded in part by the NSF QISE-NET fellowship under grant number 1747426. N.B. would also like to acknowledge Duke University for the hospitality as a majority of the work was done at Duke while N.B. was a visiting student.  

\bibliography{main}

\end{document}